\documentclass[%
reprint,
 amsmath,amssymb,
 aps,
]{revtex4-2}
\usepackage{enumitem}
\usepackage{graphicx}
\usepackage{dcolumn}
\usepackage{bm}
\usepackage{hyperref}
\usepackage{slashed}
\usepackage{subfloat}
\usepackage{multirow}
\usepackage[table]{xcolor}
\usepackage{pgfplots,subfigure}
\usepackage{orcidlink}
\usepackage{float}
\usepackage{geometry}
\usepackage[table]{xcolor} 
\usepackage{booktabs}       
\usepackage{array}          
\usepackage{hyperref}
\usepgfplotslibrary{fillbetween}
\usepackage{colortbl,xcolor} 
\usepackage{tikz}
\usetikzlibrary{arrows.meta}

\newcolumntype{C}{>{\centering\arraybackslash}m{4cm}}
\definecolor{acsblue}{RGB}{17,76,139}

\begin{document}

\fontsize{8}{9}\selectfont
\title{Optical and Thermodynamic Signatures of Lorentz Symmetry Breaking in Bumblebee AdS Black Holes}

\author{Behzad Eslam Panah}
\email{eslampanah@umz.ac.ir}
\affiliation{Department of Theoretical Physics, Faculty of Basic Sciences,
University of Mazandaran, P. O. Box 47416-95447, Babolsar, Iran.}
\affiliation{Center for Theoretical Physics, Khazar University, 41 Mehseti Str., Baku, AZ1096, Azerbaijan}


\author{Abdullah Guvendi}
\email{abdullah.guvendi@erzurum.edu.tr }
\affiliation{Department of Basic Sciences, Erzurum Technical University, 25050, Erzurum, Türkiye}
\author{Semra Gurtas Dogan}
\email{semragurtasdogan@hakkari.edu.tr (Corresponding Author)}
\affiliation{Department of Medical Imaging Techniques, Hakkari University, 30000, Hakkari, Türkiye}
\author{Omar Mustafa}
\email{omar.mustafa@emu.edu.tr }
\affiliation{Department of Physics, Eastern Mediterranean University, 99628, G. Magusa, north Cyprus, Mersin 10 - Türkiye}

\date{\today}

\begin{abstract}
{\fontsize{8}{9}\selectfont \setlength{\parindent}{0pt}
We investigate the impact of spontaneous Lorentz symmetry breaking on scalar wave propagation, null geodesics, and thermodynamic behavior of four-dimensional asymptotically AdS black holes in bumblebee gravity. \textcolor{black}{The static, spherically symmetric solutions are characterized by a dimensionless parameter $\ell > -1$ arising from the vacuum expectation value of the bumblebee vector field, which globally rescales the radial geometry.} Massless scalar fields are analyzed via the radial Klein--Gordon equation cast into a generalized Helmholtz form, yielding an effective frequency-dependent refractive index that identifies oscillatory and evanescent regions, classical turning points, and confinement induced by curvature and Lorentz violation. \textcolor{black}{In the high-frequency limit, wave propagation coincides with null geodesics, with $\ell$ controlling radial scaling and governing the geometric-optics limit.} The AdS boundary reflects waves, while the horizon acts as a one-way absorber. Thermodynamic analysis in non-extended and extended phase spaces confirms the first law and Smarr relation, with $\ell$ influencing heat capacity, free energy, and stability. \textcolor{black}{Modeling these black holes as heat engines, we construct explicit cycles and show that efficiency increases with $\ell$, leading to an upper bound imposed by $\eta \leq 1$.} Our results provide a framework connecting Lorentz violation, wave propagation, geometric optics, and AdS black hole thermodynamics in bumblebee gravity.
}
\end{abstract}


\maketitle

\tableofcontents

\section{Introduction}
\label{sec:intro}

\setlength{\parindent}{0pt}

General relativity (GR) provides an exceptionally successful theoretical framework for describing gravitational phenomena across a wide range of physical scales, including planetary motion, stellar dynamics, and the large-scale evolution of the Universe~\cite{Clifford}. Its key predictions, such as the perihelion precession of Mercury~\cite{Einstein} and gravitational lensing~\cite{Lens}, have been confirmed with high precision. Nevertheless, GR is fundamentally a classical theory, and a complete description of gravity is widely expected to require unification with quantum mechanics~\cite{QG}. Near the Planck scale, quantum fluctuations are anticipated to alter the geometric structure of spacetime, leading to deviations from classical behavior. These considerations motivate systematic investigations of quantum gravity candidates and modified theories of gravity, which can offer insights into high-energy gravitational phenomena, early-Universe cosmology, and the physics of compact astrophysical objects~\cite{QG-1,QG-2,QG-3}. In this context, spontaneous Lorentz symmetry breaking provides a consistent and phenomenologically motivated framework to study deviations from local Lorentz invariance without violating general covariance.

\vspace{0.1cm}
\setlength{\parindent}{0pt}

Lorentz invariance is a fundamental symmetry of both GR and the Standard Model of particle physics~\cite{Kibble,Smolin,C-1,C-2}, ensuring that physical laws are invariant under transformations between inertial frames. This symmetry enforces the universality of the speed of light and governs the local causal structure of spacetime. However, various approaches to quantum gravity and high-energy effective field theories suggest that Lorentz symmetry may be deformed, violated, or spontaneously broken under specific conditions~\cite{Kibble,Smolin,C-1,C-2,C-3,C-4,C-5}. Studying Lorentz symmetry breaking in gravitational contexts thus provides a theoretically consistent framework for probing departures from GR and exploring physics beyond the classical regime~\cite{Kibble,Smolin,C-1,C-2,C-3,C-4,C-5}.

\vspace{0.1cm}
\setlength{\parindent}{0pt}

Bumblebee gravity constitutes a minimal and well-motivated extension of GR in which a vector field $\mathcal{B}_\mu$ acquires a nonzero vacuum expectation value $b_\mu$, thereby inducing spontaneous Lorentz symmetry breaking~\cite{B-0,B-1,B-2,ref-paper}: \(\langle \mathcal{B}_\mu \rangle = b_\mu\). The emergence of a preferred spacetime direction modifies the gravitational dynamics by introducing corrections to the Einstein field equations and generating anisotropic contributions to spacetime curvature. At the same time, bumblebee models retain the essential geometric structure of GR, making them particularly suitable for isolating and analyzing the physical consequences of Lorentz symmetry breaking on gravitational fields, cosmological evolution, and black hole spacetimes~\cite{B-1,ref-paper}. \textcolor{black}{In recent years, the space of exact solutions in bumblebee gravity has expanded considerably. New classes of static, stationary, and axisymmetric vacuum spacetimes have been constructed, including general classifications of spherically symmetric solutions, metric-affine extensions, and exact vacuum configurations, thereby considerably enlarging the phenomenological landscape of Lorentz-violating gravity and providing new settings for investigating strong-field gravitational physics~\cite{NEW1,NEW2,NEW3,NEW4,NEW5}.}

\vspace{0.1cm}
\setlength{\parindent}{0pt}

A systematic formulation of Lorentz-violating effects in particle physics is provided by the Standard Model Extension (SME), originally developed by Colladay and Kostelecký~\cite{17,C-2}. This effective field theory incorporates additional gauge-invariant operators constructed by contracting Standard Model fields with fixed background tensors~\cite{19,20}. Extensive studies within the SME framework have led to stringent constraints on Lorentz-violating parameters across a wide range of sectors, including charge conjugation, parity, and time-reversal (CPT) symmetry violation~\cite{21,22,23,24}, CPT-odd and CPT-even gauge sectors~\cite{25,26,27,28,29}, fermionic interactions~\cite{30,31,32,33}, radiative corrections~\cite{34,35,36,37,38,39,40,41}, and photon-fermion couplings~\cite{42,43,44,45}.

\vspace{0.1cm}
\setlength{\parindent}{0pt}

\textcolor{black}{The influence of bumblebee gravity on compact gravitational systems has been investigated in considerable detail, including black holes, traversable wormholes, compact objects, and nonminimal matter couplings~\cite{ref-paper,54,55,56,57,58,hassanabadi2025ergosphere,NEW5,NEW6}.} Traversable wormhole solutions constructed in this framework satisfy the standard energy conditions for normal matter and fulfill the flare-out conditions at the throat~\cite{46}, while the bumblebee background induces nontrivial global topologies that affect gravitational lensing properties. Quasinormal modes of such wormholes have been studied using the WKB approximation for both scalar and gravitational perturbations~\cite{47}. In black hole spacetimes, Lorentz symmetry breaking modifies the Hawking temperature through purely geometric corrections, independently of the spin of the emitted particles~\cite{48}. \textcolor{black}{The bumblebee field also alters quasinormal spectra, greybody factors, and scalar perturbations of rotating black-hole spacetimes, demonstrating that Lorentz violation leaves characteristic imprints on wave propagation and stability~\cite{49,50,NEW12}.} Additional effects appear in test particle motion around slowly rotating, axially symmetric, charged black holes~\cite{51} and in the properties of black hole shadows~\cite{52,53}. \textcolor{black}{Lorentz symmetry breaking has also been shown to influence the propagation of fermionic fields. In particular, neutrino motion in bumblebee backgrounds exhibits observable modifications induced by the Lorentz-violating geometry, providing an additional phenomenological probe of these theories~\cite{NEW7}. More recently, several exact black-hole solutions have considerably broadened the phenomenology of Lorentz-violating gravity, including noncommutative bumblebee black holes, new asymptotically vacuum black-hole geometries, Taub--NUT-like configurations, phantom conformal nonlinear-electrodynamic solutions, and their corresponding thermodynamic and shadow properties~\cite{54,55,56,57,58,ref-paper,NEW8,NEW9,NEW10,NEW11,NEW12}.}

\vspace{0.1cm}
\setlength{\parindent}{0pt}

Bumblebee fields also give rise to significant cosmological consequences. Analyses of the dynamical equations governing cosmic expansion demonstrate that late-time de Sitter behavior can be recovered while placing constraints on the parameters of the spontaneous symmetry breaking potential~\cite{59}. The presence of a nonzero bumblebee background can generate cosmological anisotropies, establish preferred spatial directions, and contribute to observed features of the cosmic microwave background~\cite{60}. Constraints derived from quadrupole and octopole moments indicate that the bumblebee field may also prolong the duration of the matter-dominated era~\cite{61}. Additional cosmological implications of bumblebee gravity are discussed in Refs.~\cite{62,63,64}.

\vspace{0.1cm}
\setlength{\parindent}{0pt}

The action of bumblebee gravity generalizes the Einstein-Hilbert action by incorporating a vector field nonminimally coupled to the Ricci tensor, together with a potential enforcing a nonzero vacuum expectation value~\cite{ref-paper,B-0}:
\begin{equation}
\begin{split}
\mathcal{I} = \int_{\partial \mathcal{M}} d^4x \, \sqrt{-g} \Bigg[
& \frac{\mathcal{R} - 2 \tilde{\Lambda}}{2\kappa}
+ \frac{\xi}{2\kappa} \mathcal{B}_\mu \mathcal{B}_\nu \mathcal{R}^{\mu\nu}
- \frac{1}{4} \mathcal{B}_{\mu\nu} \mathcal{B}^{\mu\nu} \\
& - \mathcal{V}(\mathcal{B}_\mu \mathcal{B}^\mu \pm b^2)
+ \mathcal{L}_{\rm matter}
\Bigg].
\end{split}
\end{equation}
Here $\kappa = 8\pi G / c^4$ denotes the gravitational coupling, $\tilde{\Lambda}$ is the cosmological constant, $\xi$ controls the strength of the nonminimal coupling, and $\mathcal{V}$ enforces the nonzero vacuum expectation value. The effective cosmological constant generated by the Lagrange multiplier field is given by \(\Lambda = \frac{\kappa \lambda}{\xi}\).

\vspace{0.1cm}
\setlength{\parindent}{0pt}

Eslam Panah~\cite{ref-paper} demonstrated that the Lorentz-violating parameter $\ell = \xi b^2$ plays a decisive role in determining the horizon structure and thermodynamic stability of bumblebee AdS black holes, modifying their heat capacity and free energy in both non-extended and extended phase spaces and leading to a super-entropy condition for $\ell>0$ consistent with $C_P < 0$.

\vspace{0.1cm}
\setlength{\parindent}{0pt}

\begin{figure}[ht]
\centering
\includegraphics[scale=0.65]{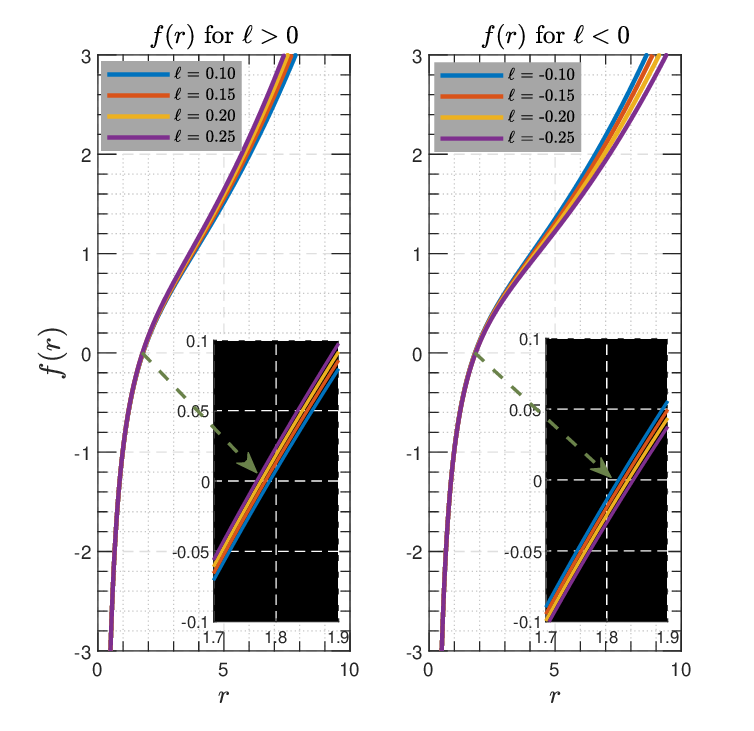}\\
\caption{\fontsize{8}{9}\selectfont Metric function $f(r)$ as a function of $r$ for different values of the Lorentz-violating parameter $\ell$ in bumblebee gravity, with $m_0=1$ and $\Lambda=-0.1$. The left panel corresponds to positive $\ell$, while the right panel corresponds to negative $\ell$. The insets present magnified views near the horizon region, showing how Lorentz symmetry breaking shifts the horizon location. For $\ell>0$, the black-hole horizon radius decreases, whereas for $\ell<0$ the horizon radius increases.}
\label{fig:0}
\end{figure}

\begin{figure}[ht]
\centering
\includegraphics[scale=0.65]{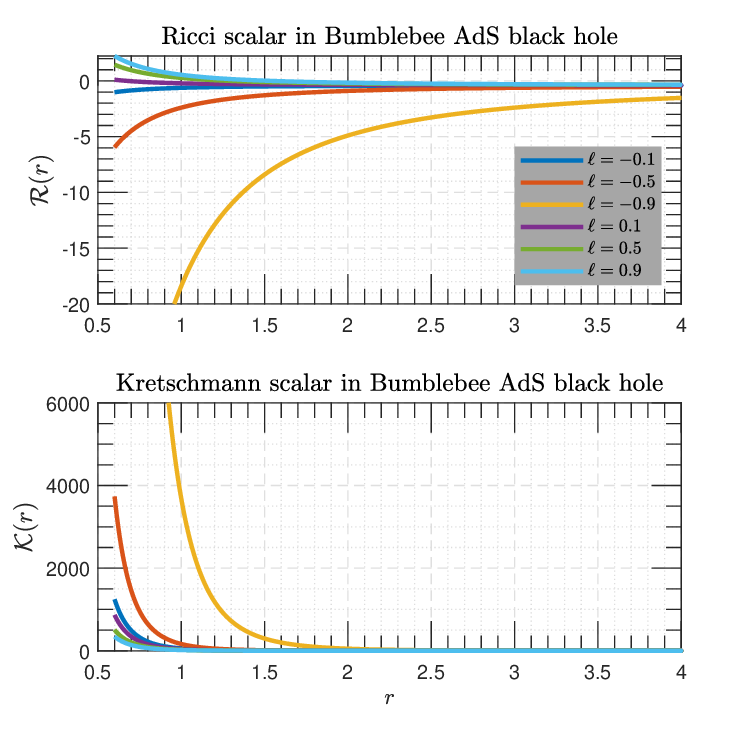}\\
\caption{\fontsize{8}{9}\selectfont Radial profiles of curvature invariants plotted as functions of the radial coordinate $r$ for the Lorentz-violating AdS black hole in bumblebee gravity with $m_0 = 1$ and $\Lambda = -0.1$. Top panel: Ricci scalar $\mathcal{R}(r)$ versus $r$. Bottom panel: Kretschmann scalar $\mathcal{K}(r) = R_{\mu\nu\rho\sigma} R^{\mu\nu\rho\sigma}$ versus $r$. Different curves correspond to distinct values of the Lorentz-violating parameter $\ell$, with $\ell > -1$ ensuring a well-defined metric signature. Negative $\ell$ enhances curvature near the origin, whereas positive $\ell$ suppresses it. At large radii, all curves approach their AdS asymptotic values, indicating that Lorentz violation predominantly affects the near-horizon and strong-field regions.}
\label{fig:curv-inv}
\end{figure}

Static, spherically symmetric solutions in bumblebee gravity generalize the Schwarzschild-(A)dS geometry~\cite{ref-paper,B-0}:
\begin{equation}
ds^2 = -f(r)\, dt^2 + \frac{1+\ell}{f(r)}\, dr^2 + r^2 \bigl(d\theta^2 + \sin^2\theta \, d\varphi^2\bigr),
\label{eq:metric}
\end{equation}
with a radial bumblebee field $\mathcal{B}_\mu = (0, b_r(r), 0, 0)$ and
\begin{equation}
f(r) = 1 - \frac{2m_0}{r} - (1+\ell)\,\frac{\Lambda}{3}\, r^2 .
\label{lapse-func}
\end{equation}
The factor $1+\ell$ rescales the effective cosmological term and consequently shifts the locations of the horizons, which are determined by the real positive roots of $f(r)=0$. For $\Lambda>0$, increasing $\ell$ enlarges the black hole event horizon while moving the cosmological horizon inward, whereas for $\Lambda<0$ the single AdS black hole horizon decreases in size as $\ell$ increases. These trends follow directly from the dependence of the roots of $f(r)$ on $\ell$ and remain significant even for $|\ell|\ll 1$. Accordingly, small deviations from $\ell=0$ generate non-negligible changes in the horizon geometry, as shown in Figure~\ref{fig:0}. Curvature invariants reveal the geometric modifications induced by the bumblebee field. The Ricci scalar is~\cite{ref-paper}
\begin{equation}
\mathcal{R} = 4\Lambda + \frac{2\ell}{(1+\ell)\, r^2},
\end{equation}
and the Kretschmann scalar reads~\cite{ref-paper}
\begin{equation}
\begin{split}
\mathcal{R}_{\alpha\beta\gamma\delta} \mathcal{R}^{\alpha\beta\gamma\delta}
&= \frac{8\Lambda^2}{3}
+ \frac{8\Lambda\, \ell}{3(1+\ell)\, r^2}
+ \frac{4\ell^2}{(1+\ell)^2\, r^4}
+ \frac{16 m_0 \ell}{(1+\ell)^2\, r^5}\\
&+ \frac{48 m_0^2}{(1+\ell)^2\, r^6}.
\end{split}
\end{equation}
These invariants diverge at \(r=0\) and remain finite elsewhere. Asymptotically,
\[
\lim_{r\to\infty} \mathcal{R} = 4\Lambda,
\qquad
\lim_{r\to\infty} \mathcal{K} = \frac{8}{3}\Lambda^2,
\]
indicating that the spacetime becomes asymptotically (A)dS for \(\Lambda>0\) (dS) or \(\Lambda<0\) (AdS)~\cite{ref-paper}. The influence of \(\ell\) is strongest near the black hole and decreases at large distances (see Figure \ref{fig:curv-inv}). The Ricci scalar exhibits an explicit $\ell$-dependent $r^{-2}$ correction superimposed on the constant AdS background value $4\Lambda$, indicating that Lorentz symmetry breaking introduces a subleading, long-range geometric correction that decays algebraically with radius. The Kretschmann scalar diverges as $r^{-6}$ near the origin, confirming the persistence of a curvature singularity whose strength is strongly modulated by $\ell$. Negative values of $\ell$ amplify the curvature invariants and signal an enhancement of the strong-field regime, whereas positive $\ell$ reduces curvature magnitudes without altering the asymptotic AdS structure.
These results demonstrate that spontaneous Lorentz violation acts as a controlled geometric deformation that reshapes local curvature while preserving the global spacetime behavior.

\vspace{0.1cm}
\setlength{\parindent}{0pt}

Massless scalar fields constitute sensitive probes of curved spacetime geometry, since their propagation directly reflects the causal structure and curvature of the underlying background~\cite{a1,a2,a3,a4,a5,a6,a7,a8,a9,a10,a11,o1,o2,o3,o4,o5,konoplya}. Scalar wave dynamics encode information about horizon properties, stability, and observational signatures of Lorentz symmetry breaking~\cite{Filho,hassan}. In bumblebee black hole backgrounds, scattering behavior, absorption cross sections, and related wave phenomena exhibit characteristic features associated with the nonzero vacuum expectation value~\cite{Zhang}. These effects are relevant for the interpretation of astrophysical observations, including gravitational waves emitted by compact objects and electromagnetic signals from accretion processes~\cite{49,Senjaya,acc}. Analog systems provide complementary platforms for investigating Lorentz-violating effects. Acoustic, optical, and condensed matter systems can emulate aspects of curved spacetime geometry, including effective horizons, wave propagation, and mode amplification~\cite{TO-1,TO-2,TO-3,TO-4,TO-5,TO-6,TO-7,TO-8,TO-9}. Bumblebee gravity provides a useful theoretical framework for drawing analogies between these systems and high-energy gravitational phenomena. In this work, we investigate massless scalar wave propagation in four-dimensional bumblebee AdS black hole spacetimes, emphasizing the role of spontaneous Lorentz symmetry breaking. By introducing an effective refractive index, we establish a direct correspondence between geometric deformations induced by the bumblebee field and wave dynamics in optical and analog systems.

\vspace{0.1cm}
\setlength{\parindent}{0pt}

The geometric modifications associated with spontaneous Lorentz symmetry breaking naturally prompt the question of how wave propagation responds to the altered spacetime structure. \textcolor{black}{In the high-frequency regime, wave propagation probes the underlying causal geometry, which makes massless scalar fields particularly suitable analytical tools.} In curved spacetimes, scalar wave dynamics encode information about null trajectories, effective potentials, and horizon structure, allowing a precise connection between field-theoretic propagation and classical null geodesic motion in the geometric-optics limit. Motivated by this perspective, we analyze scalar-wave propagation in the bumblebee AdS black hole background using an effective refractive index formulation, enabling a transparent comparison between high-frequency (eikonal) wave behavior and null geodesics of the background metric. In addition, we examine the thermodynamical properties of the bumblebee AdS black hole and investigate how Lorentz symmetry breaking influences thermal stability and phase behavior. This analysis allows us to derive physically motivated bounds on the Lorentz-violating parameter $\ell$, providing complementary constraints that arise from gravitational thermodynamics and further clarifying the role of spontaneous Lorentz violation in strong-field regimes. Despite these advances, a unified analysis connecting wave optics, null geodesics, and thermodynamic properties of Lorentz-violating AdS black holes is still lacking.

\vspace{0.1cm}
\setlength{\parindent}{0pt}

This paper is organized as follows. In Sec.~\ref{sec:scalar-wave}, we analyze the propagation of a massless scalar field in the Lorentz-violating Bumblebee AdS black hole background, derive the radial Klein-Gordon equation, and cast it into a Schr\"odinger-like form that makes the role of the Lorentz-violating parameter explicit. In Sec.~\ref{sec:ref-index}, we introduce an effective, frequency-dependent refractive index and develop an optical analogy that characterizes oscillatory and evanescent propagation regions, horizon behavior, and AdS boundary reflection. The correspondence between high-frequency scalar waves and null geodesics is established in Sec.~\ref{sec:Geodesics}, where we show that the eikonal limit reproduces the null structure of the Lorentz-violating spacetime. Thermodynamic properties of the Bumblebee AdS black hole, including stability conditions and heat-engine efficiency in extended phase space, are investigated in Sec.~\ref{sec:thermo-n}. Finally, Sec.~\ref{sec:conc} summarizes our results and discusses their implications for Lorentz symmetry breaking in strong gravitational fields. Throughout this work, geometrized units with \(c = 1 = G\) are employed.

\section{Massless Scalar Wave Propagation}
\label{sec:scalar-wave}

\setlength{\parindent}{0pt}

We consider the propagation of a massless scalar field in a spherically symmetric black hole spacetime  deformed by Lorentz symmetry breaking, as induced by a Bumblebee vector field acquiring a nonzero vacuum expectation value. The resulting geometry is characterized by the line element \eqref{eq:metric} with the Lorentz-violating parameter $\ell> -1$ encodes the influence of the bumblebee background, effectively introducing a preferred direction that subtly deforms the radial component of the metric. The lapse function $f(r)$ captures the combined effects of the black hole mass and the effective cosmological constant, and is given in equation \eqref{lapse-func}. This form clearly demonstrates that the Lorentz-violating parameter $\ell$ rescales the contribution of the cosmological term, thereby affecting the structure of spacetime and modifying the locations and properties of potential horizons.

\vspace{0.1cm}
\setlength{\parindent}{0pt}

The propagation of a massless scalar field $\Phi$ in this curved background is governed by the covariant Klein-Gordon equation, which takes the form~\cite{konoplya}
\begin{equation}
\frac{1}{\sqrt{-g}} \partial_\mu \left( \sqrt{-g} g^{\mu\nu} \partial_\nu \Phi \right) = 0.
\end{equation}
Here, the determinant $\sqrt{-g} = \sqrt{1+\ell}\, r^2 \sin\theta$ reflects the volume element of the Lorentz-violating spacetime, demonstrating explicitly how the deformation influences the radial propagation through the factor $\sqrt{1+\ell}$. Substituting the metric components, the Klein-Gordon equation reduces to a form in which the effects of spacetime curvature, spherical symmetry, and Lorentz violation are manifest in the radial and angular dependence of the scalar field. To analyze the propagation in detail, we separate variables by adopting the standard ansatz
\begin{equation}
\Phi(t,r,\theta,\varphi) = e^{-i \omega t} Y_{L m}(\theta,\varphi) R(r),
\end{equation}
where $Y_{L m}(\theta,\varphi)$ are the spherical harmonics satisfying
\begin{equation}
\Delta_{S^2} Y_{L m} = - L(L+1) Y_{L m}.
\end{equation}
This separation is exact due to the preservation of spherical symmetry in the presence of the purely radial bumblebee field. The resulting radial equation incorporates all the effects of the black hole mass, the effective cosmological constant, the angular momentum barrier, and the Lorentz-violating parameter:
\begin{equation}
\frac{1}{1+\ell} \frac{1}{r^2} \frac{d}{dr} \left( r^2 f(r) \frac{dR}{dr} \right) + \frac{\omega^2}{f(r)} R - \frac{L(L+1)}{r^2} R = 0.
\end{equation}
This equation reveals that the Lorentz-violating parameter $\ell$ effectively rescales the radial propagation, modifying the effective radial propagation weight as it moves through the curved geometry. Physically, this implies that the presence of a preferred direction due to spontaneous Lorentz symmetry breaking alters the relative balance between radial kinetic propagation and angular confinement, thereby indirectly modifying transmission and reflection properties. Expanding the radial derivative explicitly, one finds
\begin{equation}
\frac{1}{r^2} \frac{d}{dr} \left( r^2 f(r) \frac{dR}{dr} \right) = f(r) R'' + \left( \frac{2 f(r)}{r} + f'(r) \right) R',
\end{equation}
where the term $2 f(r)/r$ arises from the spherical volume element and the term $f'(r)$ encodes the gradient of the gravitational potential, including the modifications due to Lorentz violation. Consequently, the radial equation takes the more explicit form
\begin{equation}
\frac{1}{1+\ell} \left[ f(r) R'' + \left( \frac{2 f(r)}{r} + f'(r) \right) R' \right] + \frac{\omega^2}{f(r)} R - \frac{L(L+1)}{r^2} R = 0,
\end{equation}
which clearly shows the interplay between the geometric deformation, the effective potential due to curvature, and the centrifugal barrier arising from angular momentum. To cast the radial scalar equation into a canonical Schrödinger form suitable for optical analogy analysis, we perform the standard Liouville transformation by defining
\begin{equation}
R(r)=\frac{U(r)}{r\sqrt{f(r)}}
\label{eq:chi-def}
\end{equation}
This transformation absorbs the first-derivative damping term that arises due to the radial dependence of the metric function, effectively isolating the curvature-induced potential contributions. It is important to note that, although the radial operator carries an overall factor of $(1+\ell)^{-1}$ due to Lorentz symmetry breaking, this constant prefactor does not affect the elimination of the first-derivative term. The standard Liouville transformation therefore remains valid after a global rescaling of the equation, allowing the radial Klein-Gordon equation to be cast into a canonical Schr\"odinger-like form. Substituting \eqref{eq:chi-def} into the radial equation yields
\begin{equation}
U''(r)+\left[\frac{1+\ell}{f(r)^2} \omega^2 - V_0(r) \right]U(r)=0
\label{eq:Schrodinger-form}
\end{equation}
This equation effectively represents a one-dimensional Schrödinger-type equation with
\begin{equation}
V_0(r) = \frac{f''(r)}{2 f(r)} - \frac{f'(r)^2}{4 f(r)^2} + \frac{f'(r)}{r f(r)} + \frac{1+\ell}{r^2} \frac{L(L+1)}{f(r)}.
\label{eq:V0-canonical}
\end{equation}
This potential encodes the combined influence of the black hole mass, the cosmological constant, angular momentum, and Lorentz-violating deformation on the scalar field. The centrifugal term, proportional to $L(L+1)$, dominates at small $r$ for high angular momentum, while the terms involving $f'(r)$ and $f''(r)$ encode the influence of spacetime curvature and the horizon structure. The Lorentz-violating parameter $\ell$ acts as a global rescaling factor in the dominant terms of the effective potential, effectively modifying the horizon barrier, the tunneling probability, and the propagation speed of the scalar waves. The structure of $V_0(r)$ thus provides a direct bridge between the underlying geometric deformation due to spontaneous Lorentz symmetry breaking and observable features of wave propagation in such exotic black hole spacetimes.

\section{Effective Refractive Index}
\label{sec:ref-index}

\setlength{\parindent}{0pt}

\begin{figure*}[ht]
\centering
\includegraphics[scale=0.65]{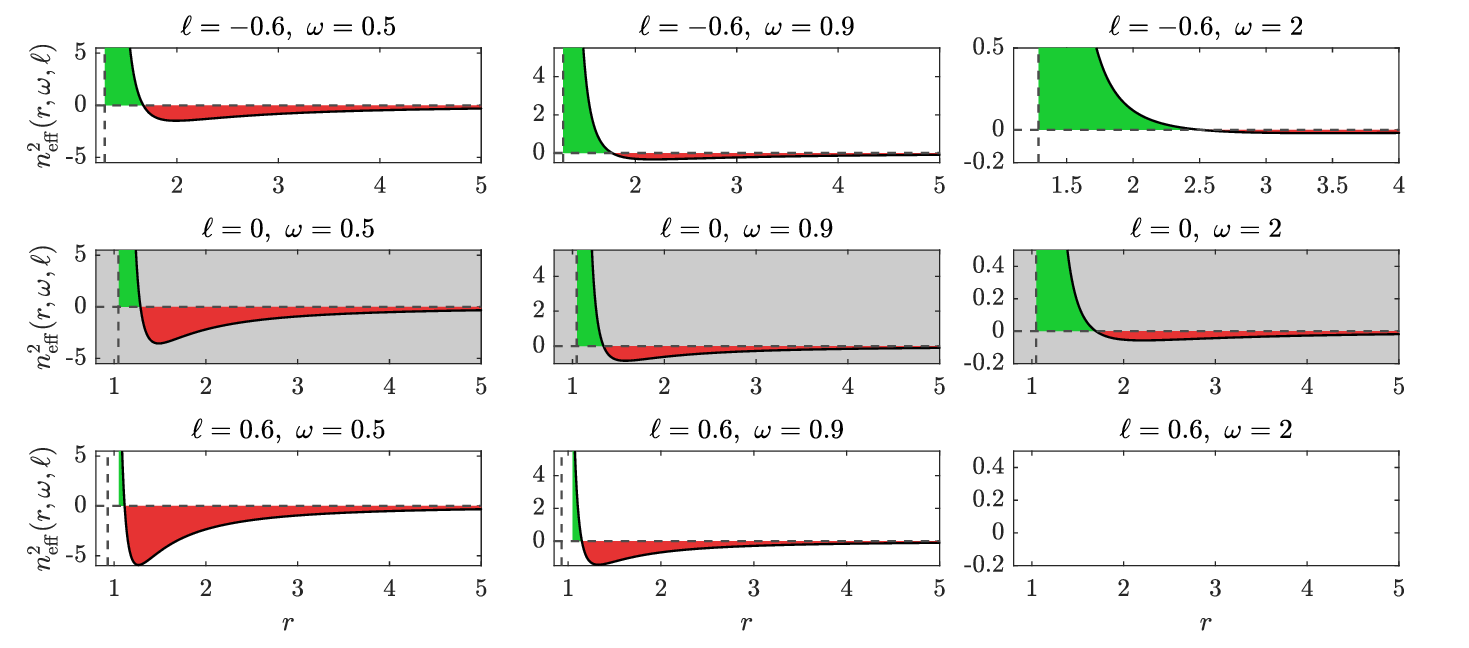}\\
\caption{\fontsize{8}{9}\selectfont \textcolor{black}{Radial profiles of $n_{\rm eff}^{2}(r,\omega,\ell)$ for scalar perturbations in the Lorentz-violating Bumblebee AdS black hole spacetime with
$M=1$, $\Lambda=-2.5$, and $L=1$. Rows correspond to $\ell=-0.6,0,0.6$, while columns correspond to
$\omega=0.5,0.9,2$. The vertical dashed line denotes the event horizon $r_h$, and the profiles are displayed in the exterior region $r_h<r\leq5$. Regions with $n_{\rm eff}^{2}>0$ ($<0$) correspond to
oscillatory (evanescent) behavior within the effective Helmholtz description. The horizontal dashed line indicates the transition condition $n_{\rm eff}^{2}=0$.}}
\label{figa}
\end{figure*}

\textcolor{black}{The radial scalar perturbation equation can be written in a generalized Helmholtz form,
\begin{equation}
U''(r)+\omega^2 n_{\rm eff}^2(r,\omega)U(r)=0 ,
\label{Helmholtz-form}
\end{equation}
where the effective refractive index is defined as
\begin{equation}
n_{\rm eff}^2(r,\omega)
=
\frac{1+\ell}{f(r)^2}
\left[
1-\frac{f(r)^2}{(1+\ell)\omega^2}V_0(r)
\right].
\label{eq:neff_complete}
\end{equation}
Here $V_0(r)$ denotes the effective potential obtained from the radial equation and includes the curvature and angular momentum contributions. The quantity $n_{\rm eff}$ should be understood as an effective quantity associated with the wave equation rather than a material refractive index. The sign of $n_{\rm eff}^2$ determines the local character of the solutions: positive values correspond to oscillatory solutions, whereas negative values indicate exponentially varying solutions. The local radial wave number is given by
\begin{equation}
k_{\rm loc}(r,\omega)=\omega n_{\rm eff}(r,\omega).
\end{equation}
Near the horizon, $f(r_h)=0$, and the metric function can be expanded as
\begin{equation}
f(r)\simeq f'(r_h)(r-r_h)
\equiv \kappa(r-r_h).
\end{equation}
Using the near-horizon behavior of the effective potential,
\begin{equation}
V_0(r)\simeq
-\frac{1}{4(r-r_h)^2}
+\mathcal{O}\!\left((r-r_h)^{-1}\right),
\end{equation}
the effective refractive index behaves as
\begin{equation}
n_{\rm eff}^2(r,\omega)
\simeq
\frac{1}{(r-r_h)^2}
\left[
\frac{1+\ell}{\kappa^2}
+\frac{1}{4\omega^2}
\right],
\qquad r\rightarrow r_h .
\end{equation}
Therefore,
\begin{equation}
k_{\rm loc}(r,\omega)
\simeq
\frac{\omega}{r-r_h}
\sqrt{
\frac{1+\ell}{\kappa^2}
+\frac{1}{4\omega^2}
}.
\end{equation}
For the allowed range $\ell>-1$, the coefficient remains positive, implying that the effective radial wave number is real in the immediate exterior near-horizon region. The corresponding horizon boundary condition, however, is determined independently from the standard near-horizon analysis of the
radial equation. The Lorentz-violating parameter modifies the coefficient of the divergence but does not change its $(r-r_h)^{-2}$ scaling. At large distances, the geometry approaches the AdS form
\begin{equation}
f(r)\simeq
-\frac{1+\ell}{3}\Lambda r^2 ,
\qquad \Lambda<0 .
\end{equation}
Consequently,
\begin{equation}
V_0(r)
\simeq
\frac{2}{r^2}
-\frac{3L(L+1)}{\Lambda r^4}
+\mathcal{O}(r^{-4}),
\end{equation}
and the refractive index becomes
\begin{equation}
n_{\rm eff}^2(r,\omega)
\simeq
\frac{9}{(1+\ell)\Lambda^2 r^4}
-\frac{2}{\omega^2 r^2}
+\frac{3L(L+1)}{\omega^2\Lambda r^4}
+\mathcal{O}(r^{-4}).
\end{equation}
For finite nonzero frequency, the leading contribution is therefore
\begin{equation}
n_{\rm eff}^2(r,\omega)
\simeq
-\frac{2}{\omega^2 r^2}
+\mathcal{O}(r^{-4}),
\qquad r\rightarrow\infty ,
\end{equation}
showing that the effective refractive index decreases algebraically in the asymptotic AdS region. The transition between oscillatory and evanescent regions is determined by
the turning-point condition
\begin{equation}
n_{\rm eff}^2(r_t,\omega)=0 ,
\end{equation}
or equivalently,
\begin{equation}
V_0(r_t)
=
\frac{(1+\ell)\omega^2}{f(r_t)^2}.
\end{equation}
The positions of these points depend on the full structure of the metric function and the effective potential. Hence, the parameter $\ell$ modifies the detailed radial distribution of propagation regions through its effect on the background geometry and on the prefactor $(1+\ell)/f(r)^2$, while the
dominant asymptotic scaling remains unchanged. In the high-frequency limit,
\begin{equation}
\omega\rightarrow\infty ,
\qquad
n_{\rm eff}^2(r,\omega)
\simeq
\frac{1+\ell}{f(r)^2},
\end{equation}
whereas in the formal low-frequency regime,
\begin{equation}
n_{\rm eff}^2(r,\omega)
\simeq
-\frac{V_0(r)}{\omega^2}.
\end{equation}
The latter limit should be interpreted cautiously since the effective Helmholtz representation becomes increasingly sensitive to the structure of the potential as $\omega$ approaches zero. This refractive-index formulation therefore provides a useful characterization of the local propagation properties of scalar perturbations in the Lorentz-violating AdS black hole background.}

\textcolor{black}{The profiles shown in Fig.~\ref{figa} demonstrate the dependence of the effective refractive index on the Lorentz-violating parameter and the wave frequency. In the near-horizon region, all configurations exhibit positive values of $n_{\rm eff}^{2}$, indicating an oscillatory character of the effective radial equation close to the horizon. Away from the horizon, the sign of $n_{\rm eff}^{2}$ is determined by the relative contributions of the frequency-dependent term and the effective potential. For lower frequencies, the contribution of $V_0(r)/\omega^2$ becomes larger, and regions with negative $n_{\rm eff}^{2}$ can appear, corresponding to evanescent behavior within the effective Helmholtz description. As the frequency increases, the potential contribution becomes suppressed, and the refractive index becomes mainly determined by the geometric factor $(1+\ell)/f(r)^2$. The parameter
$\ell$ modifies the magnitude and radial distribution of $n_{\rm eff}^{2}$ through its effect on the background geometry, while the leading near-horizon and asymptotic properties remain unchanged.}\\

\textcolor{black}{
Furthermore, the effective optical description provides a useful perspective on local wave localization. Since
\[
\lambda_{\rm eff}\sim \frac{\lambda}{|n_{\rm eff}(r,\omega)|},
\]
a large effective refractive index corresponds to a reduced local wavelength scale. The near-horizon enhancement of $|n_{\rm eff}|$ therefore suggests a possible connection between optical properties and near-horizon wave localization, which may be relevant for future imaging studies based on a full ray-tracing or wave-optics analysis.}

\section{Null Geodesics and High-Frequency Scalar-Wave Correspondence}
\label{sec:Geodesics}

\setlength{\parindent}{0pt}

\textcolor{black}{
We consider the static and spherically symmetric Lorentz-violating AdS black hole spacetime~\eqref{eq:metric}, where $\ell>-1$ characterizes the deformation induced by the bumblebee field. The existence of the Killing vectors $\partial_t$ and $\partial_\varphi$ leads to the conservation of the energy and angular momentum of null trajectories. Restricting the motion to the equatorial plane, $\theta=\pi/2$, these quantities are given by
\begin{equation}
\mathcal{E}=f(r)\frac{dt}{d\lambda},
\qquad
\mathcal{L}=r^2\frac{d\varphi}{d\lambda},
\end{equation}
where $\lambda$ denotes an affine parameter. The null condition, $ds^2=0$, gives the radial equation
\begin{equation}
\left(\frac{dr}{d\lambda}\right)^2
=
\frac{1}{1+\ell}
\left[
\mathcal{E}^2
-f(r)\frac{\mathcal{L}^2}{r^2}
\right].
\end{equation}
Hence, the factor $(1+\ell)$ modifies the radial parametrization, whereas the turning points are determined by
\begin{equation}
\mathcal{E}^2=V_{\rm geo}(r),
\qquad
V_{\rm geo}(r)=f(r)\frac{\mathcal{L}^2}{r^2}.
\end{equation}
Equivalently, the critical impact parameter is obtained from
\begin{equation}
b^2=\frac{\mathcal{L}^2}{\mathcal{E}^2}
=\frac{r_t^2}{f(r_t)} .
\end{equation}
To establish the correspondence between scalar-wave propagation and null geodesics, we introduce the Hamilton--Jacobi function $S$. The eikonal equation for massless propagation is \cite{hassanabadi2025ergosphere}
\begin{equation}
g^{\mu\nu}\partial_\mu S\,\partial_\nu S=0 ,
\end{equation}
or equivalently the Hamiltonian constraint \cite{hassanabadi2025ergosphere}
\begin{equation}
H=\frac{1}{2}g^{\mu\nu}p_\mu p_\nu=0 ,
\qquad
p_\mu=\partial_\mu S .
\end{equation}
Using the separation ansatz
\begin{equation}
S=-\mathcal{E}t+\mathcal{L}\varphi+S_r(r),
\end{equation}
the radial momentum becomes
\begin{equation}
\left(\frac{dS_r}{dr}\right)^2
=
\frac{1+\ell}{f(r)^2}
\left[
\mathcal{E}^2-V_{\rm geo}(r)
\right].
\end{equation}
This relation gives the same radial characteristic structure as the null geodesic equation, with differences arising only from the choice of radial parameterization. The same Hamilton--Jacobi equation is obtained from the scalar field equation in the geometric-optics limit. Introducing the WKB form
\begin{equation}
\Phi=A(x)e^{iS/\epsilon},
\qquad
\epsilon\rightarrow0 ,
\end{equation}
the leading-order contribution of the Klein--Gordon equation yields the massless eikonal equation above (see also \cite{Eikonal}). Therefore, the phase propagation of scalar waves follows null characteristics in the high-frequency regime. In terms of the effective Helmholtz equation~\eqref{Helmholtz-form}, the local radial wave number is
\begin{equation}
k_r^2
=
\omega^2 n_{\rm eff}^2(r,\omega)
=
\omega^2\frac{1+\ell}{f(r)^2}
\left[
1-\frac{V_0(r)}{\omega^2}
\right].
\end{equation}
For the limit, $\omega\rightarrow\infty$, the potential contribution is subleading, and therefore
\begin{equation}
k_r\simeq
\omega\frac{\sqrt{1+\ell}}{f(r)} .
\end{equation}
Identifying the wave frequency with the conserved geodesic energy,
$\mathcal{E}\simeq\omega$, gives
\begin{equation}
p_r\simeq
\frac{\mathcal{E}\sqrt{1+\ell}}{f(r)},
\end{equation}
showing the correspondence between the scalar eikonal momentum and the Hamilton--Jacobi radial momentum. The corresponding WKB phase takes the form \cite{WKB}
\begin{equation}
\Phi(r)
=
\int k_r\,dr
\simeq
\omega\sqrt{1+\ell}
\int\frac{dr}{f(r)} .
\end{equation}
Near the horizon, where \(f(r)\simeq f'(r_h)(r-r_h)\), the radial wave number behaves as
\begin{equation}
k_r\simeq
\frac{\omega\sqrt{1+\ell}}
{f'(r_h)(r-r_h)} ,
\end{equation}
reflecting the usual logarithmic phase accumulation associated with the near-horizon infinite redshift. At large radii, the asymptotic AdS behavior suppresses the radial wave number for finite frequency. In this regime, the sign of $n_{\rm eff}^2$ is determined by the balance between the geometric term and the effective potential contribution. The Lorentz-violating parameter $\ell$ affects the scalar propagation through the modified background geometry and the corresponding optical factors. In the geometric-optics limit, however, the leading scalar phase evolution follows the same characteristic structure as null geodesics. Deviations from this correspondence arise from finite-frequency corrections, including scattering effects and wave interference beyond the eikonal approximation.}

\section{Thermodynamic Properties}

\label{sec:thermo-n}

\subsection{Thermodynamical quantities and thermal stability}

\setlength{\parindent}{0pt}

In order to calculate of the Hawking temperature and the total mass of the
bumblebee AdS black hole solutions, we find the mass ($m_{0}$) by setting $%
g_{tt}=f(r)$ equal to zero (i.e. $g_{tt}=f(r)=0$), which leads to
\begin{equation}
m_{0}=\frac{\left( 3-\left( 1+\ell\right) \Lambda r_{h}^{2}\right) r_{h}}{6}.
\label{mm}
\end{equation}
The surface gravity of the bumblebee AdS black holes is %
\textcolor{black}{defined} as~\cite{ref-paper}
\begin{equation}
\kappa =\left. \frac{g_{tt}^{\prime }}{2\sqrt{-g_{tt}g_{rr}}}=\right\vert
_{r=r_{+}}=\left. \frac{f^{\prime }(r)}{2\sqrt{1+\ell}}\right\vert
_{r=r_{h}}.  \label{k}
\end{equation}
which the Hawking temperature ($T=\frac{\kappa }{2\pi }$) leads to
\begin{equation}
T=\frac{1}{4\pi\, \sqrt{1+\ell}\,r_{h}}-\frac{\sqrt{1+\ell}\, \Lambda\, r_{h}%
}{4\,\pi }.  \label{TemF(R)CPMI}
\end{equation}
Our analysis indicate that the permissible values of $\ell$ are $\ell>-1$,
which ensures a well-defined metric signature and prevents imaginary
contributions to the black hole mass and temperature. Using the area law,
the entropy of AdS black holes in bumblebee gravity is given by
\begin{equation}
S=\pi r_{h}^{2}.  \label{S}
\end{equation}
The total mass of this AdS black hole is given by applying the
Ashtekar-Magnon-Das (AMD) approach~\cite{AMDI,AMDII}
\begin{equation}
M=\frac{m_{0}}{\sqrt{1+\ell}}=\frac{\left( 3-\left( 1+\ell\right) \Lambda
r_{h}^{2}\right) r_{h}}{6\sqrt{1+\ell}},  \label{AMDMass}
\end{equation}
which, we have to consider the limit $\ell>-1$ to avoid of a singularity at $%
\ell=-1$, and imaginary value, too. It is easy to demonstrate that the
conserved quantities and thermodynamic quantities adhere to the first law of
thermodynamics as $dM=TdS$.

\vspace{0.1cm} \setlength{\parindent}{0pt}

\textcolor{black}{In extended phase space the effective cosmological constant plays
the role of a thermodynamic pressure, given by
\begin{equation}
P=\frac{-\Lambda \left( 1+\ell \right) }{8\pi }.  \label{Lambda}
\end{equation}
The Lorentz-violating parameter ($\ell $) is shown to have an effect on the
pressure, as indicated by the equation (\ref{Lambda}). Now, we re-write the
total mass of the black hole (\ref{AMDMass}) in terms of the entropy (\ref{S}%
) and the thermodynamic pressure (\ref{Lambda}) and $\ell $, which leads to
\begin{equation}
M\left( S,P,\ell \right) =\frac{\left( 3+8PS\right) \sqrt{S}}{6\pi ^{1/2}%
\sqrt{1+\ell }}.  \label{MSP}
\end{equation}
The thermodynamic volume ($V$), and the Hawking temperature ($T$) of
bumblebee AdS black holes are given by using Eq. (\ref{MSP}), which lead to
\begin{eqnarray}
V &=&\left. \frac{\partial M\left( S,P,\ell \right) }{\partial P}\right\vert
_{S,\ell }=\frac{4\pi }{3\sqrt{\left( 1+\ell \right) }}\left( \frac{S}{\pi }%
\right) ^{3/2},  \label{V1} \\
T &=&\left. \frac{\partial M\left( S,P,\ell \right) }{\partial S}\right\vert
_{P,\ell }=\frac{1+8PS}{4\sqrt{\pi S\left( 1+\ell \right) }}.  \label{Text}
\end{eqnarray}
Using the extended mass relation (Eq. (\ref{MSP})), we can formulate the
first law of thermodynamics within the context of the extended phase, which
is
\begin{equation}
dM=TdS+VdP+\Pi d\ell ,
\end{equation}where $\ell $ is a variable parameter and $\Pi $ is defined as
\begin{equation}
\Pi =\left. \frac{\partial M\left( S,P,\ell \right) }{\partial \ell}\right\vert _{S,P}=\frac{\left( 8PS-3\right) }{12\left( 1+\ell \right) ^{3/2}}\left( \frac{S}{\pi }\right) ^{1/2}. \label{LL}
\end{equation}
By considering Eqs. (\ref{MSP}), (\ref{V1}), (\ref{Text}), and (\ref{LL})
we get the Smarr relation, which is
\begin{equation}
M=2TS-2PV.
\end{equation}
Applying $M\left( S,P,\ell\right) $, we can obtain the heat capacity at constant
pressure ($C_{P}$), the heat capacity at constant volume ($C_{V}$), and the
Helmholtz free energy in the following forms
\begin{eqnarray}
C_{P} &=&\frac{T}{\left. \frac{\partial ^{2}M\left( S,P,\ell\right) }{\partial
S^{2}}\right\vert _{P}}=\frac{2S\left( 1+8PS\right) }{8PS-1},  \label{Cpext}
\\
C_{V} &=&\left. T\left( \frac{\partial S}{\partial T}\right) \right\vert
_{V}=0.  \label{Cvext} \\
F(T,S,P,\ell) &=&M\left( S,P,\ell\right) -TS=\frac{\left( 3-8PS\right) \sqrt{S}}{12%
\sqrt{\pi \left( 1+\ell \right) }},  \label{F}
\end{eqnarray}
}
The temperature (Eq. (\ref{Text})) and the Helmholtz free energy impose a
constraint on $\ell $, and it is $\ell >-1$. To find the thermodynamical
stable area for bumblebee AdS black holes, we have to respect to three
following \textcolor{black}{conditions}, simoultaneously
\begin{equation}
\left\{
\begin{array}{ccc}
T>0 & \rightarrow  & 8PS+1>0 \\
C_{P}>0 & \rightarrow  & 8PS+1>0\text{~~~\&~~~}8PS-1>0 \\
F<0 & \rightarrow  & 3-8PS<0%
\end{array}%
\right. ,
\end{equation}%
The above conditions determine a common stable area which is $8PS>3$. By
replacing the relation $P=\frac{-\Lambda (1+\ell )}{8\pi }$ into this
condition we find that the inequality $\ell >-\frac{3\pi }{\Lambda S}-1$. On
the other hand, we must to respect the limit $\ell >-1$ to avoid of a
singularity at $\ell =-1$, and imaginary value. Whereas the cosmological
constant is negative, so, a common suitable range which satisfy the
thermodynamical stable area is
\begin{equation}
\ell >-\frac{3\pi }{\Lambda S}-1,  \label{constraintL}
\end{equation}%
we can re-write the obtained constraint for the Lorentz-violating parameter
versus entropy, which leads to
\begin{equation}
S>-\frac{3\pi }{\Lambda \left( 1+\ell \right) },  \label{constriantS}
\end{equation}%
which for large value of $\ell $, the stable area increases. As a result,
the existence of the Lorentz-violating parameter ($\ell $) affects the size
of the stable area.

\subsection{Heat engine}

\setlength{\parindent}{0pt}

\begin{figure}[tbh]
\centering
\includegraphics[width=0.5\linewidth]{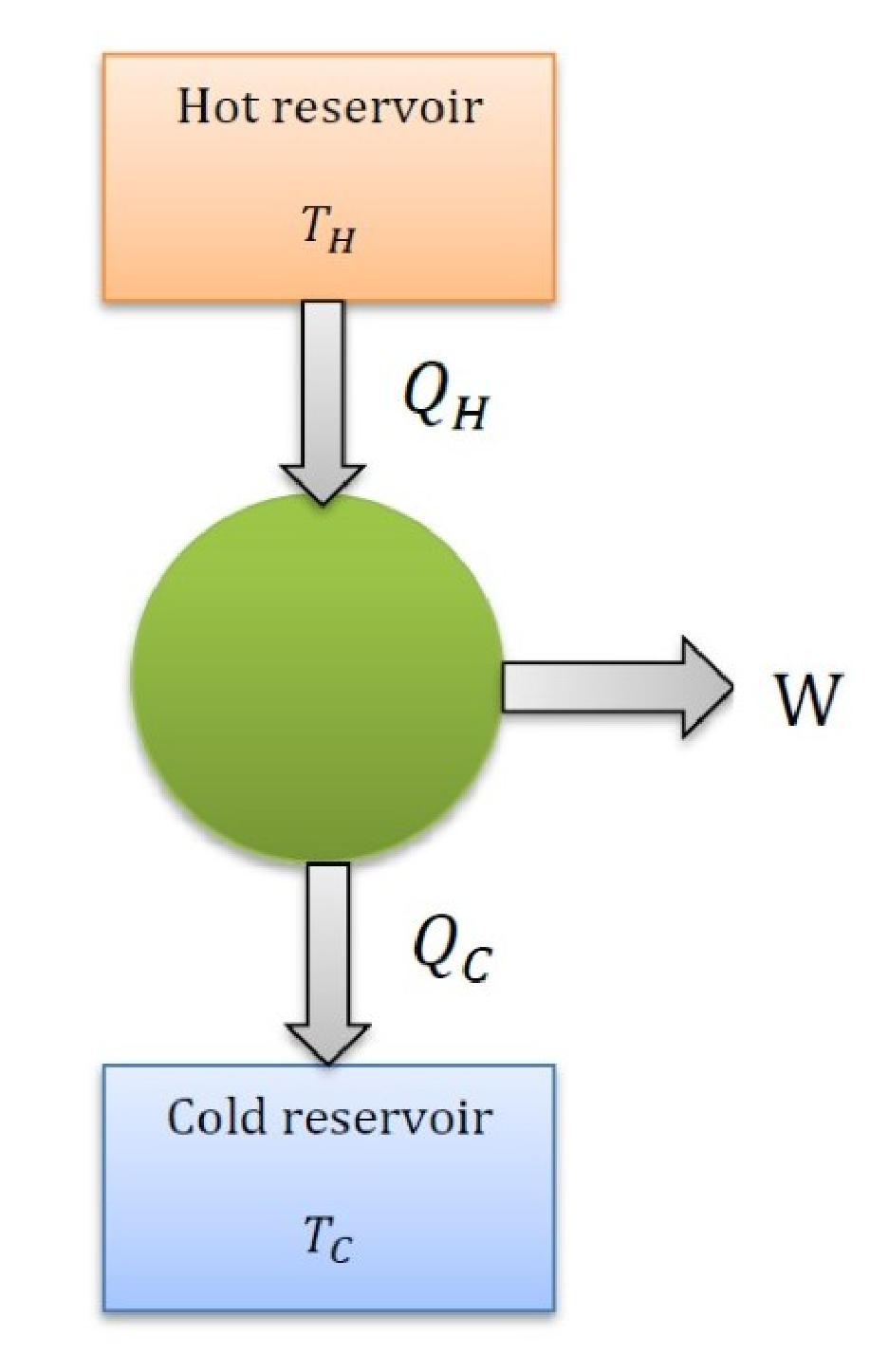}
\caption{\fontsize{8}{9}\selectfont The heat engine flows.}
\label{Fig3}
\end{figure}

\begin{figure}[tbh]
\centering
\includegraphics[width=0.4\linewidth]{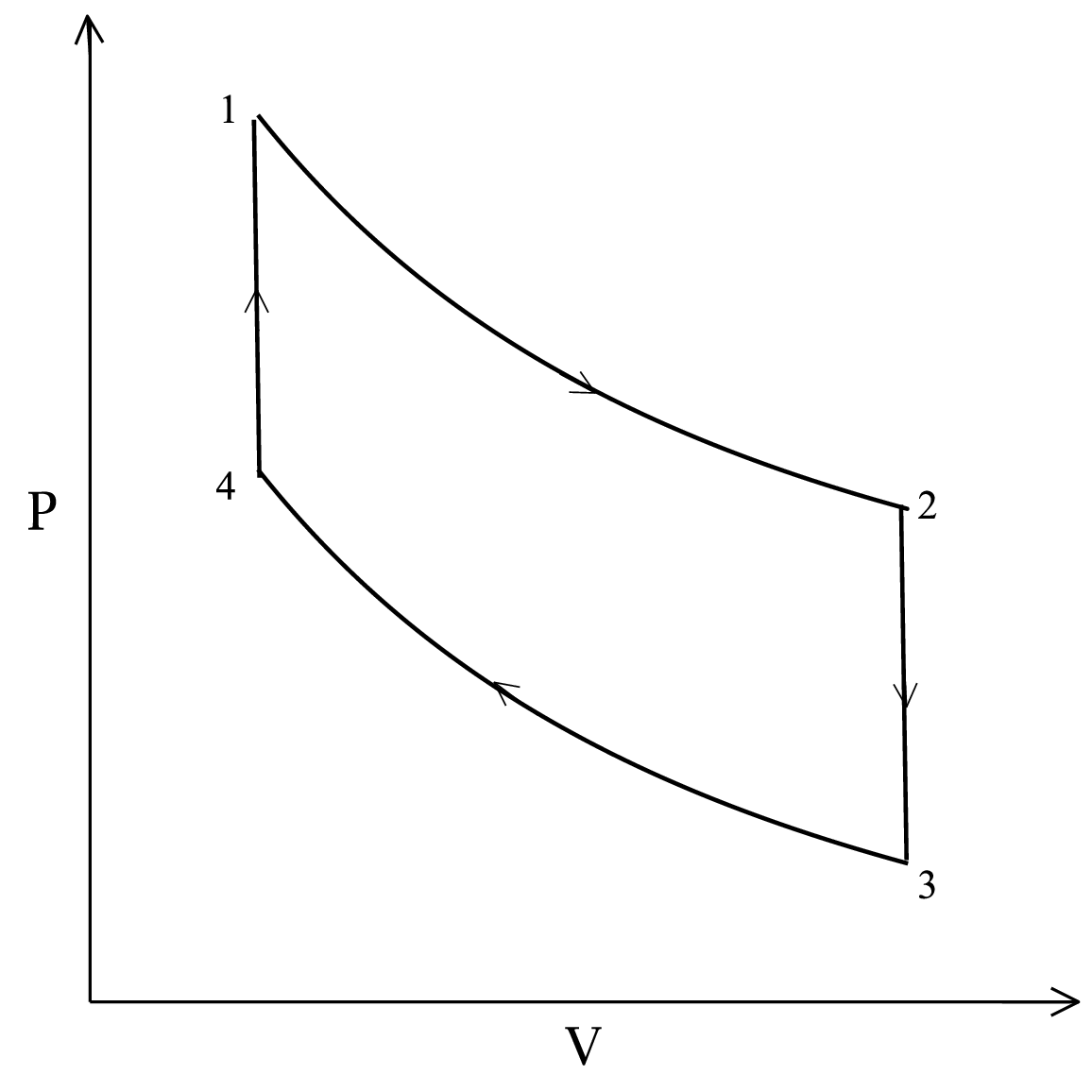}\hfil
\includegraphics[width=0.4\linewidth]{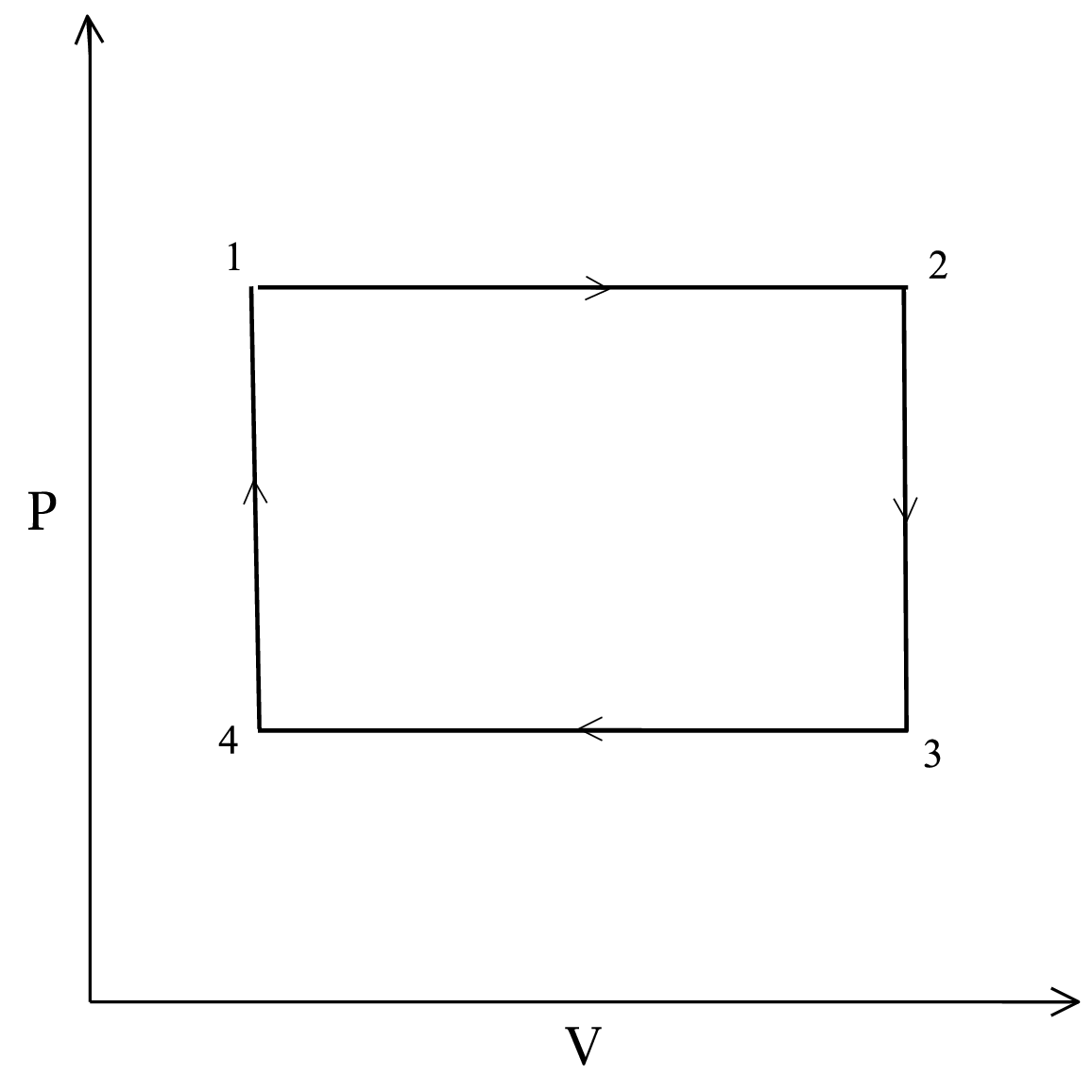}
\caption{\fontsize{8}{9}\selectfont $P$-$V$ diagram of thermodynamic cycles
for the heat engine: Left panel: Carnot engine, which for static black holes
is also a Stirling engine. Right panel: the rectangle cycle.}
\label{Fig4}
\end{figure}

Considering black holes as thermodynamic systems within the framework of
extended phase space, it is natural to interpret them as potential heat
engines. In this context, the mechanical term $PdV$ appearing in the first
law of thermodynamics allows for the computation of mechanical work, and
consequently, the efficiency of these heat engines. A heat engine can be
represented as a closed cycle in the $P-V$ diagram, operating between two
thermal reservoirs at temperatures $T_{H}$ (hot reservoir) and $T_{C}$ (cold
reservoir). During the cycle, the engine absorbs a quantity of heat $Q_{H}$
from the hot reservoir, part of which is converted into mechanical work \ $W$%
, while the remaining heat $Q_{C}${} is released to the cold reservoir (see
Figure~\ref{Fig3} for more details). The efficiency of the heat engine is
thus defined as
\begin{equation}
\eta =\frac{W}{Q_{H}}=1-\frac{Q_{C}}{Q_{H}}.  \label{Eqeff}
\end{equation}
The requirement that the engine efficiency remains below unity therefore
provides an independent and physically motivated bound on the
Lorentz-violating parameter $\ell$.
For static black holes, the thermodynamic volume $V$ and the entropy $S$ are
intrinsically interdependent variables. Consequently, the heat capacity at
constant volume ($C_{V}$) vanishes ($C_{V}=0$). This condition, $C_{V}=0$,
leads to the crucial identity where isochoric processes coincide with
adiabatic processes. As a direct result, the Carnot and Stirling heat engine
cycles become equivalent in the $P-V$ diagram (see the left panel of Figure~%
\ref{Fig4}). Given an explicit expression for the heat capacity at constant
pressure ($C_{P}$), we are motivated to define an alternative engine cycle
in the $P-V$ plane. This new cycle takes the form of a rectangle (see the
right panel of Figure~\ref{Fig4}), which is constructed from two isobaric
paths ($1\rightarrow 2$ and $3\rightarrow 4$) and two isochoric paths ($%
2\rightarrow 3$ and $4\rightarrow 1$). The net work done during this
thermodynamic cycle can then be calculated as
\begin{eqnarray}
W &=&\oint PdV=W_{1\longrightarrow 2}+W_{2\longrightarrow
3}+W_{3\longrightarrow 4}+W_{4\longrightarrow 1}  \notag \\
&&  \notag \\
&=&W_{1\longrightarrow 2}+W_{3\longrightarrow 4}=P_{1}\left(
V_{2}-V_{1}\right) +P_{4}\left( V_{4}-V_{3}\right) ,  \label{EqWo}
\end{eqnarray}%
in the above equation, the work done along paths of $2\rightarrow 3$ and $%
4\rightarrow 1$ are zero ($W_{2\longrightarrow 3}=W_{4\longrightarrow 1}=0$%
). The upper isobar will give the heat input as
\begin{equation}
Q_{H}=\int_{T_{1}}^{T_{2}}C_{P}\left( P_{1},T\right)
dT=\int_{S_{1}}^{S_{2}}C_{P}\left( P_{1},T\right) \left( \frac{\partial T}{%
\partial S}\right) dS  \notag
\end{equation}

Building upon the foundational work of Johnson in 2014, who first calculated
the efficiency of a heat engine for a black hole~\cite{Johnson}, subsequent
research has extended this analysis to a wide array of black hole
spacetimes. Utilizing Johnson's established concepts, researchers have
successfully determined the thermodynamic heat engine efficiency for
numerous systems, including: Gauss-Bonnet (GB) black holes~\cite{JohnsonII},
Born-Infeld AdS black holes~\cite{JohnsonIII}, dilatonic Born-Infeld black
holes~\cite{Bhamidipati}, rotating black holes~\cite{Hennigar}, charged AdS
black holes~\cite{Liu}, BTZ black holes~\cite{Mo,Balart2019}, polytropic
black holes~\cite{Setare}, AdS black holes in higher dimensions~\cite%
{BelhajC}, black holes in conformal gravity~\cite{Xu}, black holes in
massive gravity~\cite{Hendi}, benchmarking black holes~\cite{Chakraborty},
accelerating AdS black holes~\cite{Zhang2018}, black holes in gravity's
rainbow~\cite{EslamPanah}, charged accelerating AdS black holes~\cite%
{ZhangII}, nonlinear charged AdS black holes~\cite{Nam}, charged 4D
Einstein-Gauss-Bonnet-AdS black holes~\cite{EslamPanah2020}, charged
rotating accelerating AdS black holes~\cite{Jafarzade}, the Hayward-AdS
black hole~\cite{SGuo}, charged AdS black hole in the Gauss-Bonnet gravity
coupled with nonlinear electrodynamics~\cite{Guo2023}, and phantom AdS black
holes in noncommutative space~\cite{Hamil2025}.

\vspace{0.1cm} \setlength{\parindent}{0pt}

\begin{figure}[tbh]
\centering
\includegraphics[width=0.7\linewidth]{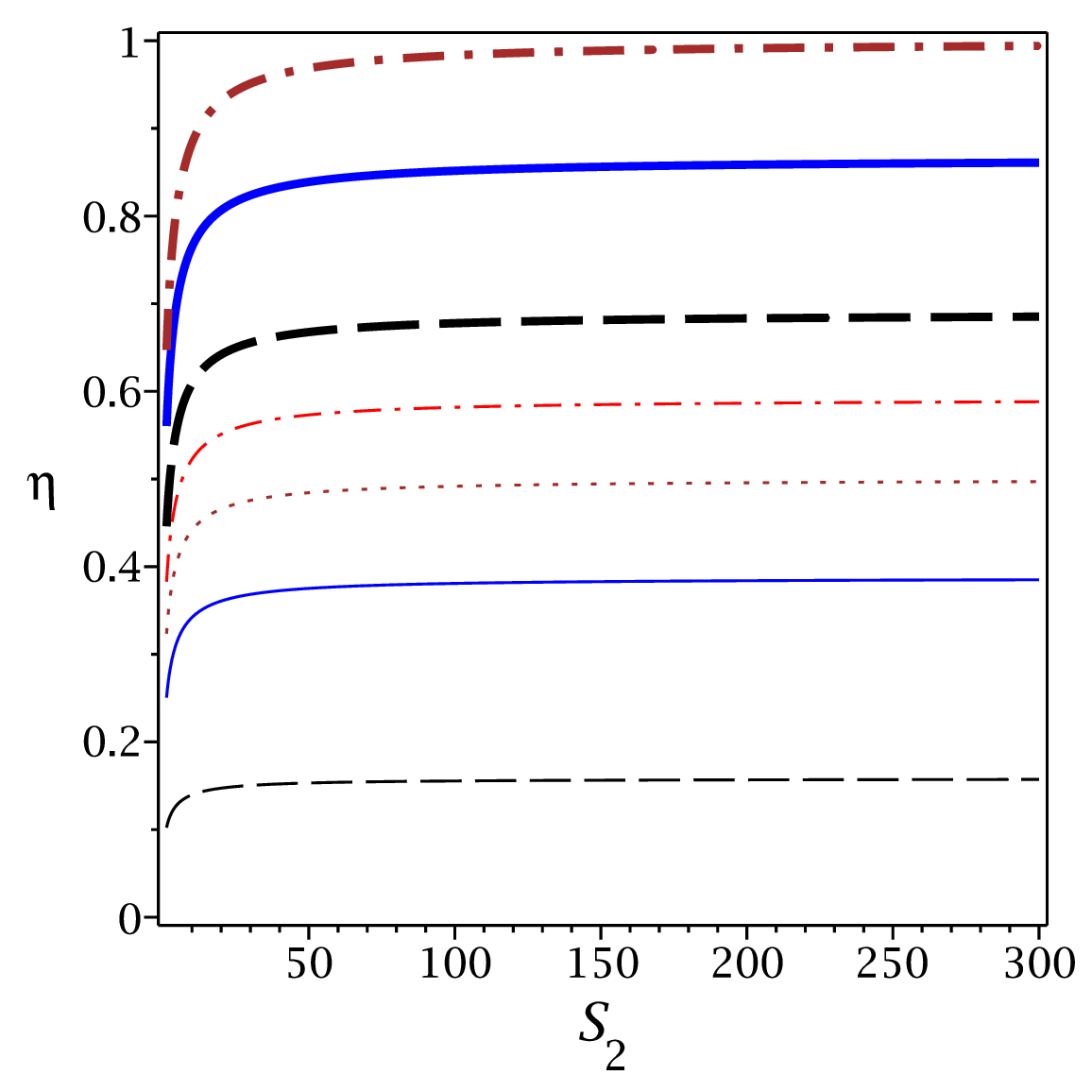}
\caption{\fontsize{8}{9}\selectfont $\protect\eta $ versus $S_{2}$ for $%
S_{1}=1$, $P_{1}=0.3$, $P_{4}=0.1$, and different values of the bumblebee
parameter ($\ell=-0.9$, $\ell=-0.4$, $\ell=0$, $\ell=0.4$, $\ell=0.9$, $%
\ell=2$, and $\ell=3$, from \textcolor{black}{down to up}).}
\label{Fig5}
\end{figure}

Using Eq. \ref{EqWo}, the useful work is obtained as
\begin{equation}
W=\frac{4\left( P_{1}-P_{4}\right) \left( \sqrt{S_{2}}-\sqrt{S_{1}}\right)
\left( S_{1}+S_{2}+\sqrt{S_{1}S_{2}}\right) }{3\sqrt{\pi }},  \label{work}
\end{equation}%
and $Q_{H}$ is calculated as
\begin{equation}
Q_{H}=\frac{8P_{1}\left( S_{2}^{3/2}-S_{1}^{3/2}\right) +3\left( \sqrt{S_{2}}%
-\sqrt{S_{1}}\right) }{6\sqrt{\pi \left( 1+\ell\right) }}.  \label{QH}
\end{equation}
Inserting Eqs. (\ref{work}) and (\ref{QH}) into Eq. (\ref{Eqeff}), one can
obtain the engine's efficiency, which is
\begin{equation}
\eta =\frac{8\left( P_{1}-P_{4}\right) \left( \sqrt{S_{2}}-\sqrt{S_{1}}%
\right) \left( S_{1}+S_{2}+\sqrt{S_{1}S_{2}}\right) \sqrt{\left(
1+\ell\right) }}{8P_{1}\left( S_{2}^{3/2}-S_{1}^{3/2}\right) +3\left( \sqrt{%
S_{2}}-\sqrt{S_{1}}\right) }.  \label{efficiency}
\end{equation}
To evaluate the effect of various parameters on the efficiency ($\eta $) of
the heat engine associated with these black holes, we plot the expression
from Eq. (\ref{efficiency}) as a function of $S_{2}$ (see Figure~\ref{Fig5}%
). Our analysis reveals that, while holding $S_{1}$, $P_{1}$, and $P_{4}$
constant, the efficiency initially rises with increasing $S_{2}$ before
asymptotically approaching a fixed value at large $S_{2}$. Furthermore, the
influence of the parameter $\ell$ on efficiency demonstrates a positive
correlation. In other words, $\eta $ increases monotonically as $\ell$
increases (as shown in Figure~\ref{Fig5})

\begin{figure}[tbh]
\centering
\includegraphics[width=0.48\linewidth]{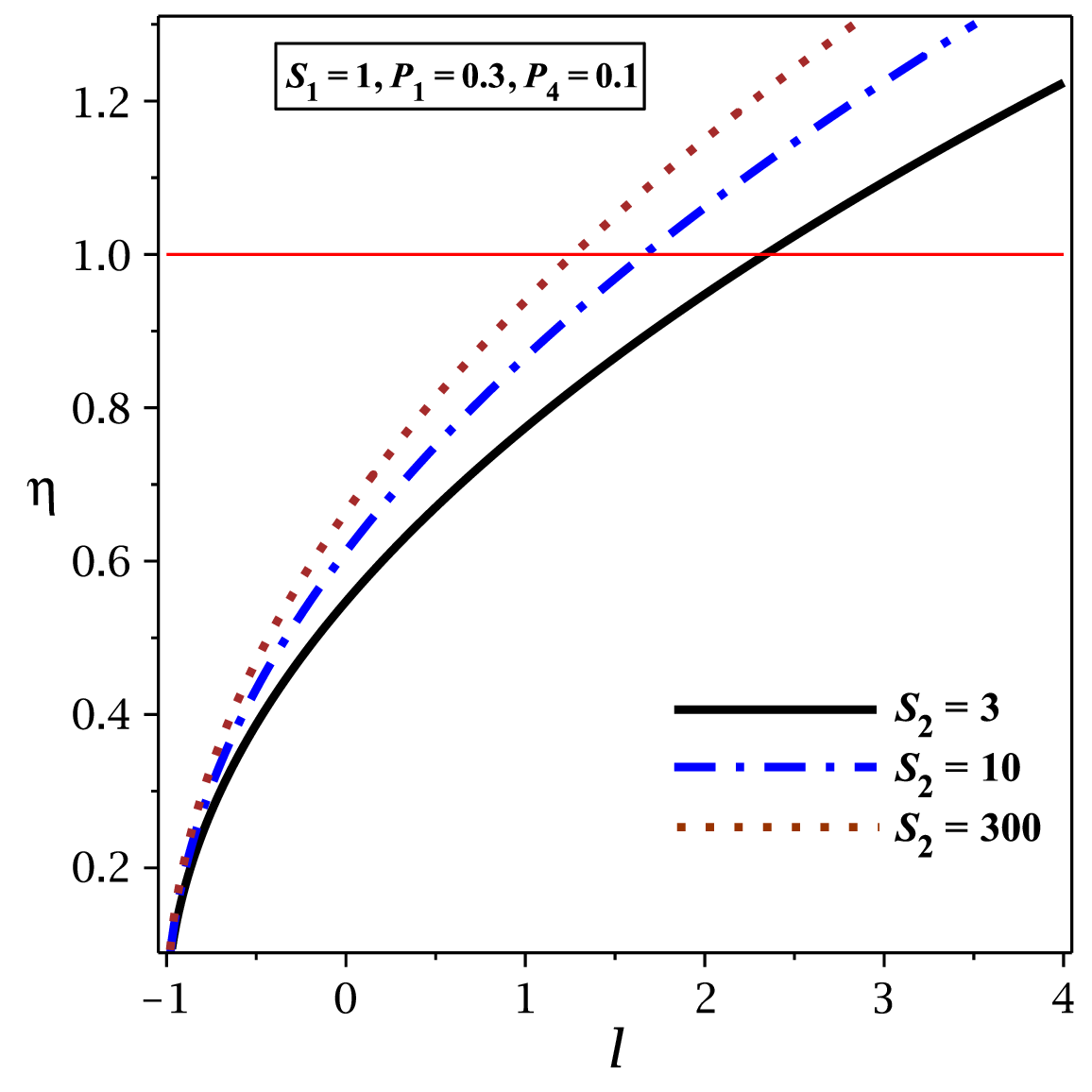}\hfil
\includegraphics[width=0.48\linewidth]{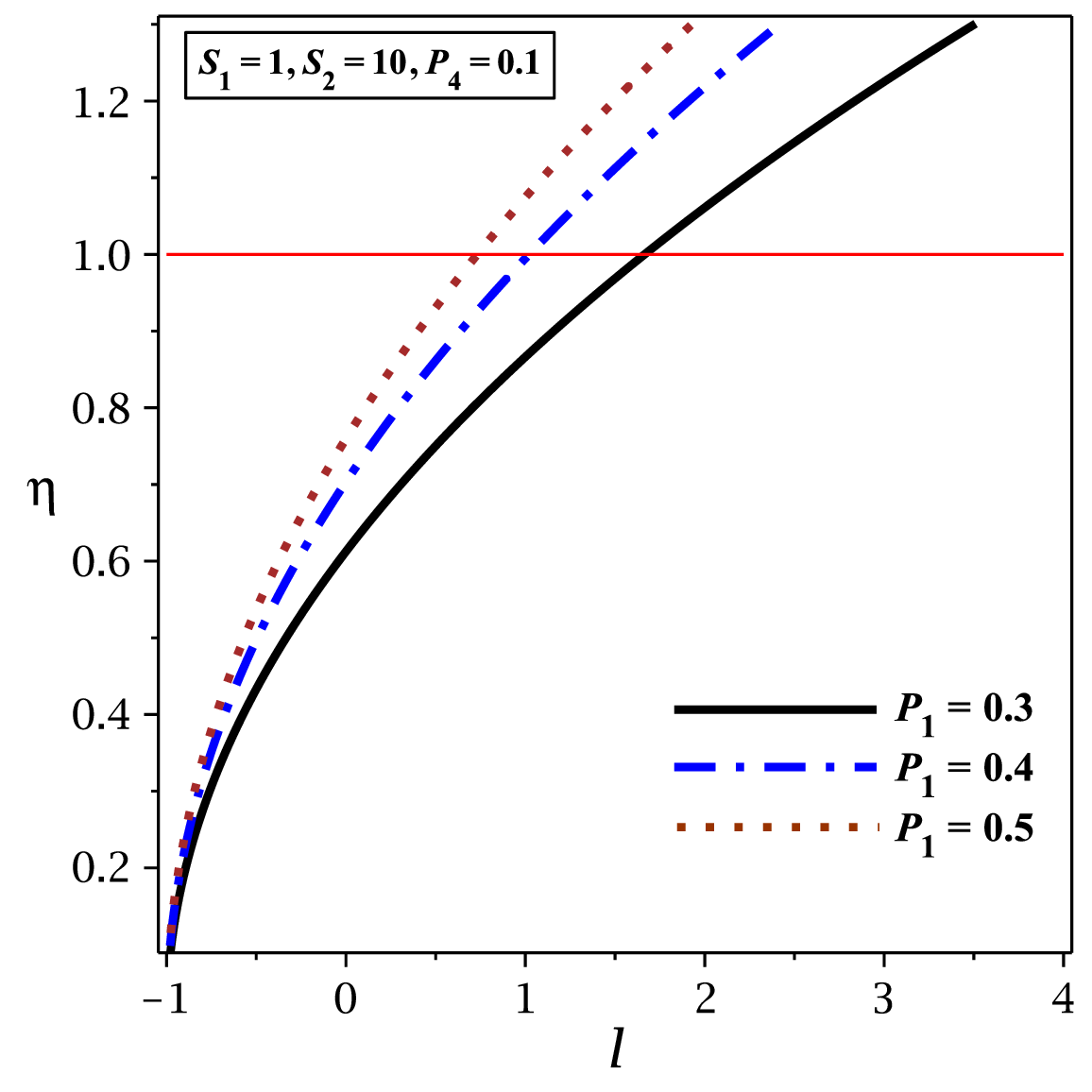}
\caption{\fontsize{8}{9}\selectfont $\protect\eta $ versus $\ell$ for
different values of parameters.}
\label{Fig6}
\end{figure}

We investigate the efficiency ($\eta $) of the heat engine associated with
the Bumblebee AdS black hole with the specific goal of determining an upper
bound for the parameter $\ell$. By imposing the physical constraint $\eta
\leq1 $, we derive an upper limit for the Lorentz-violating parameter as a
function of the pressure and entropy. This limit is described by the
following relation
\begin{equation}
\ell_{\max }=\frac{\frac{3A_{1}}{4}-\left( 4P_{1}-P_{4}\right)
A_{2}P_{4}+\left( 2P_{1}-P_{4}\right) A_{3}P_{4}+\frac{9}{64}}{%
A_{1}^{2}\left( P_{1}-P_{4}\right) ^{2}},  \label{lmax}
\end{equation}%
where $A_{1}$, $A_{2}$, and $A_{3}$ are in the following forms
\begin{eqnarray}
A_{1} &=&\left( S_{1}+S_{2}+\sqrt{S_{1}S_{2}}\right) ,  \notag \\
A_{2} &=&\left( S_{2}-S_{1}\right) \sqrt{S_{1}S_{2}},  \notag \\
A_{3} &=&\left( S_{1}^{2}+S_{2}^{2}+3S_{1}S_{2}\right) .
\end{eqnarray}
The constraint derived in Eq. (\ref{lmax}) establishes that the
Lorentz-violating parameter ($l$) must remain below a specific threshold,
namely $\ell<\ell_{\max }$. To systematically investigate how the parameters
$S_{2} $ and $P_{1}$ influence the permissible range of $\ell$, we plot the
efficiency $\eta $ as a function of $\ell$ for various fixed values of \ $%
P_{1}$ and $S_{2} $ (Figure~\ref{Fig6}). This analysis reveals that an
increase in either $S_{2}$ or $P_{1}$ leads to a reduction in the maximum
allowable value of $\ell$. Specifically, the effect of $S_{2}$ and $P_{1}$
on the permissible range of $\ell$ can be observed in the left and right
panels of Figure~\ref{Fig6}, respectively.
Overall, our investigation yields two independent constraints on the
Lorentz-violating parameter ($\ell$), derived from analyzing both the
local/global stability conditions and the physical requirement $\eta \leq 1$%
. Specifically, the lower bound for $\ell$ is established by the stability
analysis, resulting in the condition: $\ell>-\frac{3\pi }{\Lambda S}-1$.
Conversely, the upper bound for $\ell$ is determined by applying the
thermodynamic constraint $\eta \leq 1$ (see Eq. (\ref{lmax})), which ensures
that the heat engine efficiency does not exceed unity. By treating the
bumblebee AdS black holes as thermodynamical systems, we computed their
corresponding thermodynamical quantities. Subsequently, we demonstrated that
these quantities satisfy the first law of thermodynamics in both extended
and non-extended phase spaces, from which the Smarr relation was obtained.
Subsequently, we evaluated the heat capacity at constant pressure ($C_{P}$),
the Hawking temperature ($T$), and the Helmholtz free energy ($F$) in order
to comprehensively investigate the regions of local and global stability
within the extended phase space. This analysis yielded a lower bound for the
Lorentz-violating parameter $\ell$. Moreover, we found that the presence of
this parameter significantly affects the size of the stability region. Next,
by modeling these black holes as heat engines, we determined their
efficiency, which was shown to depend on $\ell$ as well as on other
thermodynamical quantities. Our results indicated that the efficiency
increases for larger values of $\ell$; however, $\ell$ is subject to an
upper bound. Indeed, by imposing the thermodynamic constraint $\eta \leq 1$,
which ensures that the heat engine efficiency cannot exceed unity, we
derived the upper limit for the Lorentz-violating parameter $\ell$.

\section{Summary and Discussions}
\label{sec:conc}

We investigated four-dimensional asymptotically AdS black holes in bumblebee gravity, focusing on the effects of spontaneous Lorentz symmetry breaking on spacetime geometry, scalar-wave propagation, null trajectories, and thermodynamic behavior. Lorentz violation is \textcolor{black}{governed by the dimensionless parameter $\ell>-1$, originating from the bumblebee vacuum configuration $\ell=\xi b^2$, which effectively rescales the radial sector of the metric while preserving stationarity and spherical symmetry}. This deformation shifts horizon locations, modifies the near-horizon geometry, and confines curvature deviations to the strong-field region, leaving the asymptotic AdS structure unchanged. All curvature invariants remain finite outside $r=0$, confirming that Lorentz-violating effects do not introduce additional singularities.

\vspace{0.1cm}

Massless scalar fields were used as probes of the Lorentz-violating background. After separation of variables and a Liouville transformation, the radial Klein-Gordon equation was reduced to a Helmholtz form with an effective refractive index $n_{\rm eff}(r,\omega)$. Oscillatory regions correspond to $n_{\rm eff}^2>0$, while evanescent regions satisfy $n_{\rm eff}^2<0$. In the high-frequency limit, the potential term becomes negligible and $n_{\rm eff}$ reduces to a geometric quantity, reproducing the null geodesic structure. \textcolor{black}{This establishes a correspondence between high-frequency scalar waves and null geodesics in the geometric-optics limit, where the phase propagation follows the Hamilton--Jacobi characteristics with Lorentz violation entering only as a rescaling of optical distances via $(1+\ell)$ without modifying the null structure of the characteristic equations.}

\vspace{0.1cm}

\textcolor{black}{Thermodynamic properties were analyzed in both non-extended and extended phase spaces. The Hawking temperature is suppressed by $\ell$, whereas the entropy obeys the area law. \textcolor{black}{In the extended phase space, the effective pressure is taken to be $P=-\frac{\Lambda (1+\ell)}{8\pi}$, which leads to a modification of the first law of thermodynamics. We also obtained the Smarr relation in this context.} Stability analysis using the heat capacity and Helmholtz free energy provides a lower bound $\ell_{\mathrm{min}}$ for positive temperature and local stability. Modeling these black holes as heat engines constrains $\ell$ further through efficiency requirements ($\eta \le 1$), yielding complementary constraints that define an upper bound $\ell_{\mathrm{max}}$. Lorentz violation is physically allowed only within this finite window, with geometric and thermodynamic consistency giving complementary limits.}

\vspace{0.1cm}

\textcolor{black}{The effective refractive index provides an optical framework in which wave propagation reproduces geodesic motion in the geometric-optics limit, thereby connecting optical, and kinematic properties of the black hole. The results show that spontaneous Lorentz symmetry breaking produces controlled deformations of black hole spacetimes that preserve causal and thermodynamic consistency. This approach can be extended to rotating black holes, higher-dimensional spacetimes, and holographic contexts, where Lorentz violation could affect quasinormal modes, scattering properties, and dual field-theory observables. The bounds on $\ell$ suggest that strong-field tests of gravity could
provide constraints on Lorentz-violating physics.}

\vspace{0.1cm}

\textcolor{black}{Thus, spontaneous Lorentz symmetry breaking in bumblebee gravity introduces a single deformation parameter that consistently modifies black hole geometry, wave propagation, null geodesics, and thermodynamic behavior in a mutually consistent manner, providing a framework for Lorentz-violating effects in strong gravitational fields and enabling extensions to more realistic and observationally relevant scenarios.}

\begin{acknowledgements}
B. Eslam Panah thanks University of Mazandaran. \textcolor{black}{The authors sincerely thank the reviewers for their valuable comments, insightful suggestions, and careful evaluation of the manuscript.}
\end{acknowledgements}

\end{document}